\documentclass[a4paper,twocolumn]{article}
\usepackage{epsf}
\normalsize
\def\bn{{\bf n}}
\def\bp{{\bf p}}
\def\bq{{\bf q}}
\def\bk{{\bf k}}

\def\xik{\xi_{\bk}}
\def\Im{{\rm Im}}
\def\Re{{\rm Re}}
\def\bra{\langle}
\def\ket{\rangle}
\def\up{\uparrow}
\def\down{\downarrow}
\def\eps{\epsilon}
\def\om{\omega}
\def\sg{\sigma}
\def\Sg{\Sigma}
\def\sgn{{\rm sgn}}
\begin{document}
%\setcounter{page}{0}
%%%%%%%%%%%%%%%%%%%%%%%%%%%%%%%%% TITLE PAGE %%%%%%%%%%%%%%%%%%%%%%%%%%%%
\title{Fermi Surface of the 2D Hubbard Model \\ 
       at Weak Coupling}
\author{Christoph J.\ Halboth and Walter Metzner \\
{\em Sektion Physik, Universit\"at M\"unchen} \\
{\em Theresienstra{\ss}e 37, D-80333 M\"unchen, Germany}}
\date{\small{October~30,~1996}}
\maketitle
%%%%%%%%%%%%%%%%%%%%%%%%%%%%%%%%% ABSTRACT %%%%%%%%%%%%%%%%%%%%%%%%%%%%%%
\renewcommand{\abstractname}{\normalsize{Abstract}}
\begin{abstract}
We calculate the interaction-induced deformation of the Fermi surface
in the two-di\-men\-sional Hubbard model within second order perturbation
theory. Close to half-filling, interactions enhance anisotropies of the
Fermi surface, but they never modify the topology of the Fermi surface in
the weak coupling regime.\\[1ex]
\mbox{{\bfseries PACS:}~05.30.Fk,~71.10.Fd,~71.18.+y}
\end{abstract}
\rule{\linewidth}{0.6pt}
%%%%%%%%%%%%%%%%%%%%%%%%%%%%% INTRODUCTION %%%%%%%%%%%%%%%%%%%%%%%%%%%%%
\section*{\normalsize{1~Introduction}}
 Since the discovery of high-temperature superconductivity, the 
structure of low-lying single-particle excitations in two-dimensional 
interacting Fermi systems has attracted much interest \cite{WYD,DAG}.
A key role is thereby played by the shape of the Fermi surface which
determines the phase space for residual scattering processes and 
thus decay rates and Fermi liquid instabilities \cite{PKC}.
In a recent Monte Carlo simulation of the two-dimensional Hubbard 
model Bulut, Scalapino and White \cite{BSW} have found that strong 
interactions may not only deform the Fermi surface of the non-interacting
reference system, but may even lead to a different topology. While the
Fermi surface of the non-interacting Hubbard model (with nearest-neighbor
hopping) is always closed around the origin in $\bk$-space for densities
$n<1$, the strongly interacting system exhibited a Fermi surface closed
around $(\pi,\pi)$ for densities close to but below half-filling.
This result raises the interesting question whether such a behavior
occurs only in a strong coupling regime. 
Zlati\'c, Schotte and Schliecker \cite{ZSS} have recently argued that the
Fermi surface topology can be changed by arbitrarily weak interactions in
the limit $n \to 1$, since in that limit arbitrarily small deformations
could lead to a topologically different shape.

 In the following we will analyze the in\-ter\-act\-ion-in\-duced 
deformation of the Fermi surface in the two-dimensional Hubbard model 
within second order perturbation theory. We show that interactions enhance
the anisotropy of the Fermi surface for densities close to half-filling, 
but they do not change the topology by deforming a $(0,0)$-centered 
surface into a $(\pi,\pi)$-centered one in the weak coupling regime. 

%% added in Version 1.1
The deformation of the Fermi surface in two-dimensional Hubbard models
has been studied already earlier by Sch\"onhammer and Gunnarsson
\cite{SG} and by Ossadnik \cite{OSS}. The aim of the former authors
was to show that Kohn-Sham Fermi surfaces are in general not
exact. The behavior of deformations in the limit $n \to 1$, however, has
not been addressed in these earlier works.
%%

%%%%%%%%%% FERMI SURFACE AND PERTURBATION EXPANSION %%%%%%%%%%%%%%%%%%%%%
\section*{\normalsize{2~Fermi Surface and Perturbation Expansion}}
 The Hubbard model Hamiltonian is given by
\begin{equation}\label{eq1}
 H = -t \sum_{\sg} \sum_{\bra j,j' \ket} 
   c^{\dag}_{j'\sg} \, c_{j\sg} \> + \>
   U \sum_j n_{j\up} \, n_{j\down}
\end{equation}
where $t$ is the nearest-neighbor hopping amplitude and $U$ the on-site
Coulomb repulsion. The operators $c^{\dag}_{j\sg}$ ($c_{j\sg}$) create
(annihilate) fermions with spin projection $\sg$ on lattice site $j$ 
and $n_{j\sg} = c^{\dag}_{j\sg} c_{j\sg}$.

\begin{figure}
%\picplace{6cm}
{\centering\rule{0cm}{6cm}\epsfbox{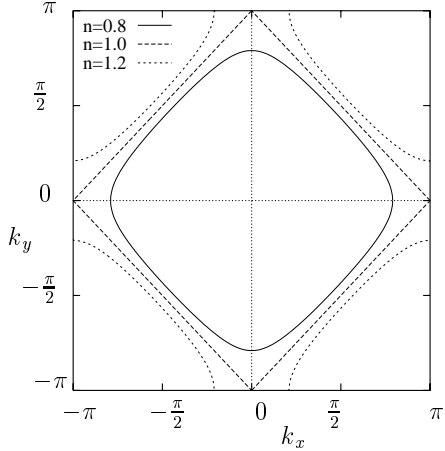}\\}
\caption{The Fermi surface of the non-interacting system ($U=0$) for 
        various densities $n$.\label{fig1}}
\end{figure}

 The non-interacting band-structure for nearest-neighbor hopping on
a square lattice is 
\begin{equation}\label{eq2}
 \eps^0_{\bk} = -2t \, (\cos k_x + \cos k_y)
\end{equation}
This leads to a strictly convex Fermi surface centered around
$\bk = (0,0)$ for densities $n<1$, to a diamond shaped surface at 
half-filling ($n\!=\!1$), and a $(\pi,\pi)$-centered Fermi surface for
densities $n>1$ (see Fig.\ \ref{fig1}). 

%%%%%%%%%% Fig.1 %%%%%%%%%%%%
%%%%%%%%%%%%%%%%%%%%%%%%%%%%%
 The Fermi surface of the interacting system can be obtained from
the one-particle Green function
\begin{equation}\label{eq3}
 G(\bk,\om) = 
   {1 \over \om - (\eps^0_{\bk} - \mu) - \Sg(\bk,\om)}
\end{equation}
where $\Sg(\bk,\om)$ is the self-energy. The chemical potential $\mu$ 
controls the average particle density $n$.
The equation 
\begin{equation}\label{eq4}
 \Re \, G^{-1}(\bk,\om) = \xik - (\eps^0_{\bk} - \mu) -
   \Re \, \Sg(\bk,\xik) = 0 
\end{equation}
determines the energy $\xi_{\bk} \equiv \eps_{\bk} - \mu$ of coherent 
single-particle excitations (quasi-particles) in the interacting
system \cite{AGD}.
The Fermi surface is the set of those points in $\bk$-space where the
excitation energy vanishes, i.e.\ $\xi_{\bk_F} = 0$, and thus
determined by
\begin{equation}\label{eq5}
 \eps^0_{\bk_F} - \mu \> + \> \Re\,\Sg(\bk_F,0) = 0 
\end{equation}
According to the Luttinger theorem \cite{LUT}, the volume enclosed by
the Fermi surface is related to the particle density by
\begin{equation}\label{eq6}
 n = 2 \int {d^2 k \over (2\pi)^d} \> 
   \Theta(\mu-\eps_{\bk})
\end{equation}

 We now calculate the interaction induced deformation of the Fermi
surface to second order in the coupling strength $U$. 
The particle density is kept fixed by a suitable choice of an
interaction-dependent chemical potential $\mu = \mu(n,U) = \mu_0 + 
\delta\mu$, where $\mu_0 = \mu_0(n)$ is the chemical potential
corresponding to density $n$ at $U=0$. We denote the Fermi wave 
vectors of the non-interacting system by $\bk_{F0}$ and the deformation
vectors by $\delta\bk_F = \bk_F - \bk_{F0}$.

 To first order in $U$, the self-energy is a real constant:
$\Sg_1(\bk,\om) = U n/2$. To keep the density fixed, the chemical
potential has to be shifted accordingly by $\delta\mu_1 = U n/2$,
which cancels $\Sg_1$ completely in $G$.

\begin{figure}
%\picplace{4cm}
{\centering\rule{0cm}{4cm}\epsfbox{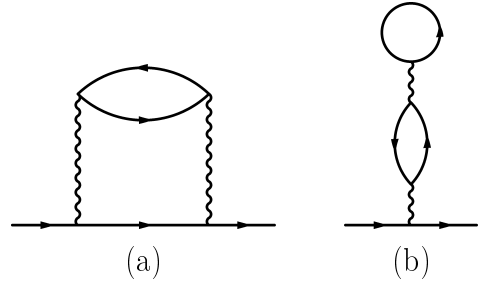}\\}
\caption{Feynman diagrams contributing to the second order 
self-energy.\label{fig2}}
\end{figure}

 To second order in $U$, two Feynman diagrams contribute to the
self-energy of the Hubbard model (see Fig.\ \ref{fig2}). 
%%%%%%%%%%% Fig. 2 %%%%%%%%%%
%%%%%%%%%%%%%%%%%%%%%%%%%%%%%
The second one is a
real constant which is completely cancelled by a corresponding shift
of the chemical potential.\footnote
{We note that this diagram is "anomalous" in the sense of Kohn and
Luttinger \cite{KL}, i.e.\ it vanishes if the thermodynamic limit is
taken strictly at $T=0$, while it yields a finite contribution if the
zero temperature limit is taken after the thermodynamic limit.}
The first diagram, however, leads to a $\bk$-dependent contribution and
generates a Fermi surface deformation.
Expanding Eqs.\ (\ref{eq5}) and (\ref{eq6}) with $\bk_F = \bk_{F0} + 
\delta\bk_F$ and $\mu = \mu_0 + \delta\mu$ in powers of $U$ and comparing 
the second order terms, one obtains the relation
\begin{equation}\label{eq7}
 \nabla\eps^0_{\bk_{F0}} \cdot \delta\bk_{F2} = 
   \delta\mu_2 - \Sg_2(\bk_{F0},0)
\end{equation}
for the deformation of the Fermi surface to second order in $U$, where
the chemical potential shift is given by the Fermi surface average
\begin{equation}\label{eq8}
 \delta\mu_2 = {\int d^2k \: \delta(\eps^0_{\bk} \!-\! \mu) \,
   \Sg_2(\bk,0) \over
   \int d^2k \: \delta(\eps^0_{\bk} \!-\! \mu)}
\end{equation}
Note that $\Sg_2$ is real at zero frequency. Of course there are many 
ways to define a map $\bk_{F0} \mapsto \bk_F$ between non-interacting 
and interacting Fermi surfaces. A natural choice is $\delta\bk_F =
\delta k_F \, \bn_{\bk_{F0}}$, where $\bn_{\bk_{F0}}$ is
a unit vector normal to the non-interacting Fermi surface in 
$\bk_{F0}$. Eq.\ (\ref{eq7}) then determines the modulus of the shift as
\begin{equation}\label{eq9}
  \delta k_{F2} = \delta k_{F2}(\bk_{F0}) = 
   {\delta\mu_2 - \Sg_2(\bk_{F0},0) \over v^0_{\bk_{F0}}}
\end{equation}
where $v^0_{\bk_{F0}} = |\nabla\eps^0_{\bk_{F0}}|$ is the Fermi
velocity of the non-interacting system.
%% added in Version 1.1:
(See also the equivalent expression (4) in \cite{SG}) 

 The second order self-energy has not yet been evaluated by
purely analytical means. 
Using the spectral representation of the non-interacting propagator
$G_0$, one can write the imaginary part of the self-energy 
contribution associated with the first Feynman diagram in Fig.\ \ref{fig2}
as
\begin{eqnarray}\label{eq10}
 \Im\Sg_{2a}(\bk,\om)& =&  - {\sgn(\om) \over \pi} \, U^2 \,
   \int {d^2 q \over (2\pi)^2} \, \int_0^{\om} d\nu \,\nonumber \\
 & &  \Im\Pi_0(\bq,\nu) \, \Im G_0(\bk\!-\!\bq,\om\!-\!\nu)
\end{eqnarray}
where the imaginary part of the non-interacting polarisation bubble
is given by
\begin{eqnarray}\label{eq11}
\Im\Pi_0(\bq,\nu) &=& {\sgn(\nu) \over \pi} \,
   \int {d^2 p \over (2\pi)^2} \, \int_{-\nu}^0 d\om' \,\qquad\qquad
 \nonumber \\
 & &\Im G_0(\bp,\om') \, \Im G_0(\bp+\bq,\om'+\nu)          
\end{eqnarray}
and $\Im G_0(\bk,\om) = 
-\pi \, \sgn(\om) \, \delta\big[\om - (\eps_{\bk}^0 - \mu_0) \big]$. 
Performing the integrals over $\om'$, $p_x$ and $q_x$ in (\ref{eq10}) 
and (\ref{eq11}) analytically (eliminating thus three
$\delta$-functions),
one obtains the expression
\begin{eqnarray}\label{eq12}
\Im\Sg_{2a}(\bk,\om)= - \sgn(\om) \, {U^2 \over 64\pi^3 t^2} \, 
   \int\! dq_y \int\! dp_y \int_0^{\om}\! d\nu&&\nonumber \\ 
\qquad \qquad \sum_{q_x^0,p_x^0} 
   \left[ {\Theta(\mu_0\!-\!\eps_{\bp}^0) - 
    \Theta(\mu_0\!-\!\eps_{\bp+\bp}^0)
   \over \sqrt{(1 \!-\! f^2)(4\sin^2(q_x/2) - g^2)}} 
   \right]_{p_x = p_x^0 \atop q_x = q_x^0}&&
\end{eqnarray}
with the functions 
\begin{equation}\label{eq13}
 \left. \begin{array}{rcl}
   f & = & (\nu - \om - \mu_0)/2t - \cos(k_y-q_y) \\
   g & = & \nu/2t + \cos(p_y+q_y) - \cos(p_y)
   \end{array} \right.
\end{equation}
The summation variables $q_x^0$ and $p_x^0$ in (\ref{eq12}) are the roots of 
the set of equations
\begin{equation}\label{eq14}
 \left. \begin{array}{rcl}
   \cos(k_x \!-\! q_x)                & = & f \\ 
   2\sin(p_x \!+\! q_x/2) \sin(q_x/2) & = & g
   \end{array} \right.
\end{equation}
The remaining three-fold integral is easily computed numerically 
(e.g.\ via a standard Monte-Carlo routine on a work-station).
Note that the representation in (\ref{eq12}) is particularly suitable for a
high resolution of the low-energy limit $\om \to 0$, since the
integration region shrinks with $\om$. 
The full self-energy function can be reconstructed from its imaginary
part $\Im\Sg_{2a}$ by a simple Hilbert transform
\begin{equation}\label{eq15}
 \Sg_{2a}(\bk,\om) = - \,\pi^{-1} \int_{-\infty}^{\infty} d\om'
   {|\Im\Sg_{2a}(\bk,\om')| \over 
    \om - \om' + i0^+ \sgn(\om)}
\end{equation}
The constant contribution $\Sg_{2b}$ from the second diagram in 
Fig.\ \ref{fig2} need not be calculated since it is completely cancelled by
a corresponding shift $\delta\mu_{2b}$ of the chemical potential.
Our numerical results for the second order self-energy agree with 
those published recently by Zlati\'c et al.\ \cite{ZSS}, who computed 
$\Sg_{2a}$ via a sequence of fast Fourier transforms. 
They also agree with earlier results by Schweitzer and Czycholl 
\cite{SC} and by Ossadnik \cite{OSS}.

%%%%%%%%%%% RESULTS FOR THE FERMI SURFACE DEFORMATION %%%%%%%%%%%%%%%%%%%%
\section*{\normalsize{3~Results for the Fermi Surface Deformation}}
\begin{figure}
%\picplace{7cm}
{\centering\rule{0cm}{7cm}\epsfbox{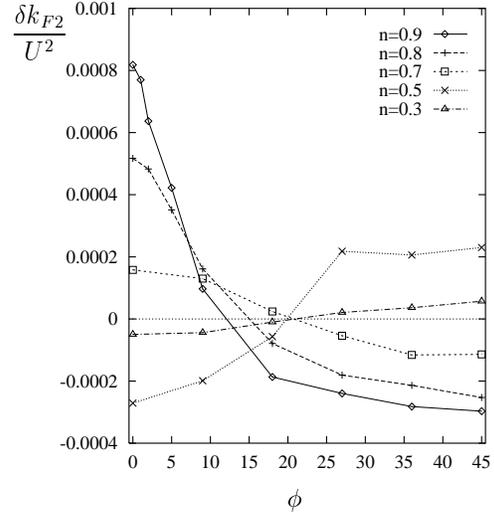}\\}
\caption{Shifts of the Fermi surface $\delta k_{F2}/U^2$
 as a function of the angle $\phi$ for various densities $n$.\label{fig3}}
\end{figure}

 To discuss explicit results for the Fermi surface deformation we
introduce polar coordinates in $\bk$-space. Points on the Fermi
surface of the non-interacting system are thus specified by an angle
$\phi$. Due to the discrete symmetries of the square lattice it is
sufficient to consider angles between $0$ and $45$ degrees.
In Fig.\ \ref{fig3} we show results for $\delta k_{F2}/U^2$ as a function of
$\phi$ for various densities $n$.
%%%%%%%%%%% Fig 3 %%%%%%%%%%%
%%%%%%%%%%%%%%%%%%%%%%%%%%%%%
Here and in the following we set the hopping amplitude $t=1$.
At low densities $n < 0.6$ weak interactions tend to compensate
anisotropies of the non-interacting Fermi surface while at densities
$0.7 < n < 1$ anisotropies are further enhanced by interactions,
as observed already earlier in unpublished work by Ossadnik
\cite{OSS}.
The diamond shaped Fermi surface at half-filling is of course not
affected at all by interactions, as a consequence of the 
particle-hole symmetry of the Hubbard model (\ref{eq1}). More generally,
particle-hole symmetry maps the Fermi surface at density $n$
onto the surface for density $2-n$ by a simple $(\pi,\pi)$-shift
in $\bk$-space (the diamond at half-filling is thereby mapped onto
itself). 

\begin{figure}
%\picplace{7cm}
{\centering\rule{0cm}{7cm}\epsfbox{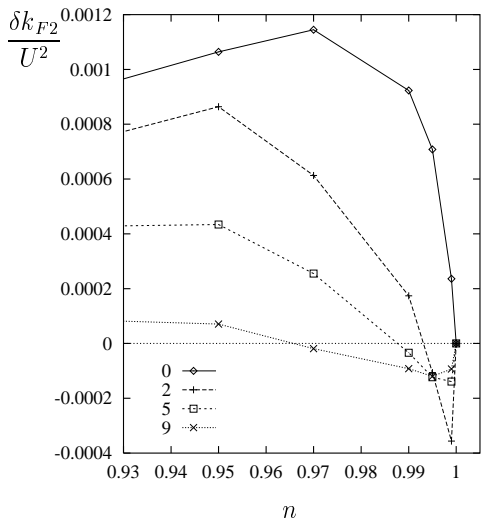}\\}
\caption{$\delta k_{F2}/U^2$ as a function of $n$ (close
to half-filling) for various small angles $\phi$.\label{fig4}}
\end{figure}

\begin{figure}
%\picplace{7cm}
{\centering\rule{0cm}{7cm}\epsfbox{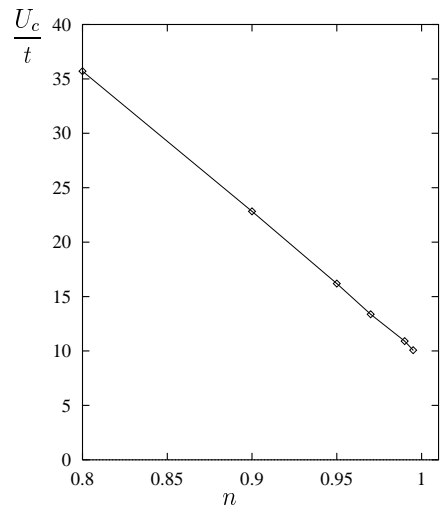}\\}
\caption{Critical coupling $U_c(n)$ required to change the
 topology of the Fermi surface.\label{fig5}}
\end{figure}

 Let us now clarify whether weak interactions can modify the
Fermi surface topology at densities close to half-filling. 
The first Fermi point that may reach the Brillouin zone boundary 
(and thus introduce a different topology) upon increasing $U$ is
obviously the one at
$\phi = 0$, since it is closer to the zone boundary than any other
Fermi point already in the non-interacting system and in addition
$\delta k_F(\phi)$ is maximal for $\phi = 0$ (if $n > 0.7$).
Quantitative information on the Fermi surface deformation in the
"critical" regime $n \to 1$ and $\phi \to 0$ is provided in Fig.\ 
\ref{fig4},
where we have plotted $\delta k_{F2}/U^2$ as a function of density 
for various small angles $\phi$. 
%%%%%%%%%%%% Fig 4 %%%%%%%%%%
%%%%%%%%%%%%%%%%%%%%%%%%%%%%%
Within numerical accuracy, $\delta k_{F2}(\phi)$ tends to $0$ for
$n \to 1$ for all $\phi$, as expected from particle-hole symmetry
and continuity. 
Finally, in Fig.\ \ref{fig5} we show the critical coupling strength $U_c(n)$ 
that is required to make $\bk_F = \bk_{F0} + \delta\bk_{F2}$ reach
the Brillouin zone boundary (at the point $(\pi,0)$). 
Close to half-filling,
$U_c(n)$ behaves linearly as a function of density and extrapolates
to a rather big {\em finite}\/ value in the limit $n \to 1$. 
Hence, interactions do not modify the Fermi surface topology of the 
two-dimensional Hubbard model within the perturbatively controlled 
weak coupling regime at any density. 

%%%%%%%%%%%% Fig 5 %%%%%%%%%%
%%%%%%%%%%%%%%%%%%%%%%%%%%%%%
 Zlati\'c et al.\ have calculated the critical coupling strength
from second order perturbation theory for $n = 0.97$ (only), where
they obtained a smaller $U_c$ (by a factor of about $2$) than we did,
although our results for the self-energy agree. 
The discrepancy arises because these authors have determined the new
Fermi surface by directly solving Eq.\ (\ref{eq5}) with the second order 
self-energy, while we have expanded $\delta k_F$ to second order in 
$U$, which is the order we really control. In the small $U$ limit both
procedures yield the same shift to order $U^2$, but quantitative 
differences arise for finite $U$. The qualitative result that $U_c(n)$
remains finite in the limit $n \to 1$ is thereby not affected.

\vspace{1ex}
 In summary, we have calculated the in\-ter\-act\-ion-in\-duced 
deformation of the Fermi surface in the two-dimensional Hubbard model 
within second order perturbation theory. 
Close to half-filling, interactions enhance anisotropies of the
Fermi surface, but they never modify the topology of the Fermi surface in
the weak coupling regime.

%%%%%%%%%%%%%%%%% ACKNOWLEDGEMENT %%%%%%%%%%%%%%%%%%%%%%%%%%%%%%%%%%%%%%%%
\vspace{2ex}
\noindent{\em Acknowledgement}
We are very grateful to Dieter \mbox{Vollhardt} for numerous valuable 
discussions. 

%%%%%%%%%%%%%%%%%%%%%%%%% REFERENCES %%%%%%%%%%%%%%%%%%%%%%%%%%%%%%%%%%%%%
%% added in Version 1.1
\newpage
\enlargethispage*{2cm}
\renewcommand{\refname}{\normalsize{References}}

\end{document}